\documentclass[a4paper,12pt]{article}
\pdfoutput=1
\usepackage{amsmath}
\usepackage{amsfonts}
\usepackage{amssymb}
\usepackage{amstext}
\usepackage{graphicx}
\usepackage[dvipsnames]{xcolor}
\usepackage{epsfig}%
\title{\centerline \bf The nature of cosmological metric perturbations in presence of gravitational particle production}
\bigskip
\author{Kaushik Bhattacharya$^\dagger$, Anirban Chatterjee$^\ddagger$, Saddam Hussain$^\$$
\thanks{$^\dagger$kaushikb@iitk.ac.in, $^\ddagger$anirbanc@iitk.ac.in, $^\$$msaddam@iitk.ac.in}
\\
\normalsize
Department of Physics, Indian Institute of Technology, Kanpur\\ 
\normalsize
Kanpur 208016, India
}
\begin{document}
\maketitle
%%%%%%%%%%%%%%%%%%%%%%%%%%%%%
\begin{abstract}
  The present paper tries to answer the question: Can a de Sitter
  phase in presence of radiation be a competitor of the standard
  inflationary paradigm for the early universe? This kind of a de
  Sitter phase can exist in cosmological models where gravitational
  particle production takes place. To address the issue the metric
  perturbations in the de Sitter phase in presence of radiation must
  be known. The evolution of metric perturbations are explicitly
  calculated in the paper. It is seen that the
    evolution of scalar and vector perturbations are considerably
    different from standard inflationary models. These differences
    arise due to the particle production mechanism. The scalar
    perturbation power spectrum grows exponentially at small length
    scales. However, one cannot
    uniquely specify the scale at which this exponential growth starts
    because of the dependence of the power spectrum on the initial
    perturbation value. The time slice on which the initial
    perturbation starts is arbitrary. The arbitrariness of the initial
    time slice is related to the problem of setting the initial
    condition on the perturbations in the present model.  The paper
  briefly opines on vector and tensor perturbations in the de Sitter
  space filled with radiation. It is seen that quantization of the
  perturbation modes in the present model is considerably
  difficult. One has to modify the present model to provide a
  consistent scheme of quantization of the perturbation modes.
  
\end{abstract}
%%%%%%%%%%%%%%%%%%%%%%%%%%%%%
\section{Introduction}

In warm inflation models \cite{Berera:1995ie, Hall:2003zp,
  Matsuda:2009eq, Visinelli:2014qla} the universe enters a de Sitter
phase in presence of radiation. The radiation in the de Sitter phase
(strictly speaking a quasi de Sitter phase) arises from the decay of
the scalar inflaton field. The inflaton has thermal fluctuations and
these thermal fluctuations act as source to the density fluctuations
in the warm inflationary scenario.  In this paper we revisit the idea of a de
Sitter space in presence of radiation, but this radiation does not
arise from the decay of any inflaton field. We assume that the very
early universe was filled with radiation, as conventional cosmological
models assume. A part of this radiation also arises from gravitational
particle production in the de Sitter phase. Parker's original work
\cite{Parker:1969au} showed that, in the cosmological backdrop, it
is impossible to produce massless particles in a radiation dominated
universe where the scale-factor of the universe $a(t) \propto
t^{1/2}$. In the case of super cool inflation, Ford has also
calculated the energy density due to the created particles
\cite{Ford:1986sy}. Later it has been shown that one can think of
gravitational production of massless particles, in a universe filled
with radiation, if $a(t)$ is not proportional to $t^{1/2}$. If the
universe, in the presence of radiation, is in a de Sitter phase then
one can actually think of gravitational production of massless
particles \cite{Lima:2012cm}. Authors have previously worked on de
Sitter phase of the universe in presence of gravitational particle
production \cite{Gunzig:1997tk, Abramo:1996ip}. In this paper we
present the cosmological metric perturbation of the universe passing
through a de Sitter phase in presence of radiation following the
phenomenological model of particle creation as proposed by Prigogine,
Geheniau, Gunzig and Nardone in
Ref.~\cite{Prigogine:1989zz}. Gravitational particle creation in open
systems was studied in detail in Ref.~\cite{Prigogine:1989zz}.  

As particle creation can happen in a dynamical spacetime as the
Friedmann-Lemaitre-Robertson-Walker (FLRW) spacetime, the produced
particles can slowdown the rate at which energy density gets diluted
in an expanding universe. For a long time people have been working
phenomenologically on big bang models of the universe with particle
creation \cite{Alcaniz:1999hu, Wichoski:1997hx}. Previous workers have assumed
various forms of the particle creation rate as in
Ref.~\cite{Pan:2018ibu, Paliathanasis:2016dhu}. The slowdown of the
rate of the energy density of particles can be taken care of by a
negative pressure component arising purely due to the creation of
particles.  Thermodynamics of such a cosmological model has been
studied in detail in Ref.~\cite{Lima:2014hda}. Various models of
cosmological particle creation have also been used by many authors to
model the late development of the universe \cite{Baranov:2019qbc,
  Lima:2015xpa, Steigman:2008bc}.

The negative pressure arising in models with gravitational particle
production can produce a de Sitter space in the early universe if the
rate of particle production in the early universe was high. If the
universe is filled with radiation then one can have a de Sitter phase
in presence of radiation. The cosmological models for such interesting
spaces were discussed in detail by previous authors in
Ref.~\cite{Lima:2012cm}. Can this kind of a de Sitter space in
presence of radiation, in the early phase of the universe,  be
a good competitor for the traditional super cool inflationary models
where the de Sitter space originates due to the potential energy of
one scalar field \cite{Guth:1980zm, Kolb:1993rv, Linde:1983gd}? We have tried to address this important question in the present article by studying the dynamics of metric perturbations in de Sitter space in presence of radiation where gravitational particle production takes place.
In the traditional models of inflation the inflaton field produces a de
Sitter expansion and the perturbation of the inflaton field in
conjunction with other metric perturbations produce the
inhomogeneities present in the early universe. In the present case
although we have a de Sitter phase in the early universe, the scenario
lacks the inflaton perturbations. Only metric perturbations and the
accompanying perturbations in the fluid parameters are present in our
present model of accelerated expansion of the early Universe. The scalar and vector power spectrum  in the present model are very different from corresponding power spectra in the standard inflationary models. The tensor perturbation result remains identical to the standard inflationary result at linear
order where scalar and tensor perturbations decouple.
However in higher orders the tensor perturbation power spectrum is affected by the modified scalar perturbation results and consequently the tensor perturbation power spectrum, in general, will also be different from the form it has in standard inflationary theories.

In Ref.~\cite{Zimdahl:1993yu} the authors tried to formulate a similar
problem but with a different aim. In this work the authors used a
different strategy to analyze the gauge invariant cosmological
dynamics in presence of gravitational particle production.  In the
above reference the authors do not define any slow-roll parameters and
consequently do not interpret the results in the light of a slow-roll
process. The authors of Ref.~\cite{Zimdahl:1993yu} work out their
result using adiabatic perturbation on comoving hypersurfaces by
choosing the gauge where the fluid velocity potential vanishes. In the
present paper we give a complete description of the initial de
Sitter phase arising out of a universe filled with radiation and show how
this de Sitter space dynamics can be modelled as a slow-roll process.
In contrast to the above referred work the techniques used to analyze
cosmological perturbations in the present paper are new and we think
these techniques can now be refined in the near future to generalize
our method of solving cosmological dynamics in open systems. We have
presented a method by which one can compare the nature of
perturbations in de Sitter phase in presence of radiation with simple
inflationary model predictions.

One of the important observations of the present work is related to the seed perturbations or the initial value of the perturbations. Treated as a purely classical theory one has to assume the presence of the seed perturbations without any proper justification for their presence. On the other hand if one wants to quantize the scalar perturbations then the quantization procedure meets stiff challenge because of the ultraviolet instability of the theory. In presence of ultraviolet instability (related to small scale perturbation modes) it becomes difficult to define a Bunch-Davies kind of vacuum. In the standard Bunch-Davies vacua all the modes of perturbation remains stable at an earlier (conformal) time, and the initial conditions are specified in that early phase. In the absence of such a scheme the issue about the initial conditions become problematic as the low (length) scale or high $k$ modes turns out to be non-perturbative. As a result of this the power spectrum in the subhorizon limit starts to grow steeply and the exact value of $k$ for which the power spectrum starts to become steep depends upon the initial conditions applied on an arbitrary time slice. In standard inflationary theories this problem does not arise.  One probable way to address this issue is related to the assumption of a pre-de Sitter phase where the perturbation modes are well behaved at all length scales. In this paper we discuss the problem related to the initial conditions briefly in the concluding section. The proper solution of the initial value problem for the perturbations require more focussed research in this area in the future.

The material in the paper is organized in the following way. In the next section we present the relevant calculations and results required to understand the background (unperturbed) cosmology accompanied by gravitational particle production. In section \ref{dS} we present the details about the de Sitter phase in presence of radiation. Section \ref{sp} presents the results related to scalar cosmological perturbations in the de Sitter model in presence of radiation. 
In section \ref{vt} we briefly present the results related to vector and tensor perturbations in such models. The next section concludes the paper by summarizing the results obtained in it.
%%%%%%%%%%%%%%%%%%%%%%%%%%%%%%%%%%%%%%%%%%%%%%%
\section{The cosmological model with gravitational particle production}
%%%%%%%%%%%%%%%%%%%%%%%%%%%%%%%%%%%%%%%%%
In the present case we work with spatially flat FLRW metric
\begin{eqnarray}
ds^2 = dt^2 - a^2(t)\left[dr^2 + r^2 d\theta^2 + \sin^2 \theta d\phi^2 \right]\,,
\label{flrw}
\end{eqnarray}
where $a(t)$ is the dimensionless scale-factor. The universe is
pervaded by a fluid whose elements move with 4-velocity $u^\mu$ and
the 4-velocity is normalized as $u^\mu u_\mu=1$.  For the background
evolution $\bar{u}^\mu=(1,{\bf 0})$. In our notation all the barred
quantities will stand for background values. The fluid 4-velocity is
$u^\mu =\bar{u}^\mu + \delta u^\mu$, where $\delta u^\mu$ is the fluid
4-velocity perturbation. Unbarred quantities include the background value and the perturbative
fluctuations. Here we are talking about metric perturbations.  The
energy-momentum tensor of the fluid is
\begin{eqnarray}
{T}_{\rm f}^{\mu \nu} = (\rho + P)u^\mu u^\nu - P g^{\mu \nu}\,,
\label{tf}
\end{eqnarray}
where $\rho$ and $P$ are the energy density and pressure of the fluid. Here we assume the fluid to be a barotropic one with an equation of state
\begin{eqnarray}
P=\omega \rho\,,
\label{eos}
\end{eqnarray}
where $\omega$ is a constant. The fluid is made up of particles whose number may increase with time. The particle flow vector is defined as:
\begin{eqnarray}
N^\mu = n u^\mu\,,
\label{pfv}
\end{eqnarray}
where $n$ is the particle number density and is given by $n=N/V$ where the number of particles present in a physical volume $V=a^3$ is $N$. Due to matter creation there arises a pressure which gives rise to a new energy-momentum tensor as \cite{Prigogine:1989zz, Calvao:1991wg, Pan:2016jli}
\begin{eqnarray}
T^{\mu \nu}_{\rm c}=P_c(u^\mu u^\nu - g^{\mu \nu})\,,
\label{tmnc}
\end{eqnarray}
where $P_c$ is called the particle creation pressure. In the present paper the cosmological system is
  assumed to be an open thermodynamic system. The particle creation
  process makes the system open. This is a well known approach which
  started with the work of Prigogine in 1989
  \cite{Prigogine:1989zz}. Later on many authors have used this
  approach. In this approach the cosmological system is receiving
  particles from some source. It is assumed that the source is
  generating the particles gravitationally. The exact method in which the
  source creates the particles is not so important for the
  analysis. For the above mentioned reason the cosmological system is
  assumed to be an open thermodynamic system which is continuously fed
  with new particles.  The exact purpose for introducing the creation
  pressure is presented in appendix \ref{appn1}. The Einstein
equation in presence of particle creation is then given by
\begin{eqnarray}
G^{\mu \nu}=\kappa(T_{\rm f}^{\mu \nu} + T^{\mu \nu}_{\rm c})\,,
\label{efe}
\end{eqnarray}
where $\kappa=8\pi G$. Since the two fluids exchange energy-momentum we have
\begin{eqnarray}
D_\mu(T_{\rm f}^{\mu \nu} + T^{\mu \nu}_{\rm c})=0\,,
\label{emcc}
\end{eqnarray}
where the result of the action of the covariant derivative on a second rank contravariant tensor is $D_\mu T^{\mu \nu}=\partial_\mu T^{\mu \nu} + \Gamma^\mu_{\mu \alpha} T^{\alpha \nu} + \Gamma^\nu_{\mu \alpha} T^{\mu \alpha}$, and in our convention
$$\Gamma^\mu_{\alpha \beta}=\frac12 g^{\mu \sigma}(\partial_\beta g_{\alpha \sigma} + \partial_\alpha g_{\beta \sigma} - \partial_\sigma g_{\alpha \beta})\,.$$
In the present case the energy-momentum conservation equation for the unperturbed fluid is
\begin{eqnarray}
\dot{\bar{\rho}} + 3H(\bar{\rho} + \bar{P} + \bar{P}_c)=0\,.
\label{emc}
\end{eqnarray}
Here $H$ is the Hubble parameter defined as $H\equiv\dot{a}/a$. As
particles are being produced due to cosmological expansion, we must have a conservation equation as \cite{Pan:2016jli}
\begin{eqnarray}
D_\mu N^\mu = n \Gamma
\label{pncc}
\end{eqnarray}
where $\Gamma$ stands for the rate of change of particle number in a
physical volume $V$ and $n$ is the number of particles per unit
physical volume. The above relation is assumed to be true for
cosmological models with particle production although we do not know
the value of $\Gamma$ precisely. We assume throughout our work that
$\Gamma \ge 0$. 

We will see later that the creation rate can be estimated by using a
basic principle and phenomenological methods. As because expansion of
the universe is at the root of gravitational particle production we
assume that the particle creation rate is dependent on some parameter
which specifies the expansion of the universe. The only parameter in
spatially flat cosmologies which carries the information about
expansion of the universe is the Hubble parameter. We assume
$\Gamma=\Gamma(H)$ such that it vanishes when expansion of the
universe stops. Under a metric perturbation the expansion rate of the
universe remains the same as that of the background FLRW spacetime and
consequently metric perturbations do not affect the value of
$\Gamma$. In our model the fluctuations in metric variables, energy
density and number density all become locally spacetime dependent but
the particle production rate remains homogeneous and isotropic.  As we
assume that the particle production rate $\Gamma$ depends only on the
property of the background evolution we have
\begin{eqnarray}
\delta \Gamma=0\,,
\label{dgama}  
\end{eqnarray}
where $\delta \Gamma$ arises due to metric perturbations. Henceforth we will only work with an unbarred $\Gamma$ (as $H$ or $a$) whereas the other variables will have their metric fluctuations.

From the form of Eq.~(\ref{pncc}) one can write
\begin{eqnarray}
D_\mu \bar{N}^\mu = \dot{\bar{n}} + \bar{n}\bar{\Gamma}^\mu_{\mu
  0} = \dot{\bar{n}} + 3H\bar{n} = \bar{n}\Gamma\,.
\label{cpncc}
\end{eqnarray}
From the above equation we can write
$$\frac{\dot{\bar{n}}}{\bar{n}}=\Gamma-3H\,.$$ These equations
hold true for the unperturbed variables. Next we write down the
general thermodynamic relation in the fluid.

The energy conservation equation can be written as
\begin{eqnarray}
Td{s}=d\left(\frac{\rho}{n}\right)+ Pd\left(\frac{1}{n}\right)\,,
\label{1stl}
\end{eqnarray}
where $T$ is the temperature of the fluid and $s$ is the entropy per particle, or the specific entropy. Assuming adiabatic particle creation where specific entropy is conserved, i.e. when  $\dot{s}=0$, the unperturbed variables satisfy
\begin{eqnarray}
0 =\dot{\bar{\rho}} - \frac{\dot{\bar{n}}}{\bar{n}}(\bar{\rho} + \bar{P})=\dot{\bar{\rho}}
+3H\left(1-\frac{{\Gamma}}{3H}\right)(\bar{\rho} + \bar{P})\,.
\label{nec}
\end{eqnarray}
Using the above result and the relation in Eq.~(\ref{emc}) we get the expression for the unperturbed creation pressure as
\begin{eqnarray}
\bar{P}_c=-\frac{{\Gamma}}{3H}(\bar{\rho} + \bar{P})=-\frac{{\Gamma}}{3H}(1+\omega)\bar{\rho}\,.
\label{crepi}
\end{eqnarray}
As long as $\omega>-1$, $\bar{\rho}>0$ and ${\Gamma}>0$ we will
have $\bar{P}_c<0$. The above equation is like an equation of state (EOS) connecting $\bar{P}_c$ and $\bar{\rho}$ and ${\Gamma}$. As in cosmology the EOS of barotropic fluid does not depend upon orders of perturbation we assume the above EOS to be valid in general to all orders of metric perturbation and write
\begin{eqnarray}
P_c=-\frac{\Gamma}{3H}(\rho + P)=-\frac{\Gamma}{3H}(1+\omega)\rho\,,
\label{crep}
\end{eqnarray}
where up to first order $P_c=\bar{P}_c + \delta P_c$ and $\rho=\bar{\rho} + \delta \rho$. We will work with $\delta P_c$ later when we introduce metric perturbation. This calculation shows that the unperturbed pressure of the fluid due to particle creation is negative although $\delta P_c$ may not be smaller than zero.

From Eq.~(\ref{nec}) it is seen that
$\dot{\bar{\rho}} + 3H\bar{\rho}
(1+\omega)= {\Gamma} \bar{\rho}(1+\omega)$ producing
$$\frac{\dot{\bar{\rho}}}{\bar{\rho}}= (1+\omega)({\Gamma} - 3H)\,.$$
Comparing this result with the expression of $\dot{\bar{n}}/\bar{n}$ we can write
\begin{eqnarray}
(1+\omega)\frac{\dot{\bar{n}}}{\bar{n}}=\frac{\dot{\bar{\rho}}}{\bar{\rho}}\,,
\label{nre}
\end{eqnarray}
which has a solution as \cite{Lima:1995xz}
\begin{eqnarray}
\bar{n}=n_0 \left(\frac{\bar{\rho}}{\rho_0}\right)^{1/(1+\omega)}\,,
\label{nrr}
\end{eqnarray}
where $n_0$ and $\rho_0$ are the values of the variables at some given instant of cosmic time.

The temperature evolution due to cosmological expansion is affected by the particle production mechanism\cite{Calvao:1991wg, Weinberg:1971mx}. Assuming $\bar{T},\bar{n}$ to be the basic thermodynamic variables and taking $\bar{\rho}=\bar{\rho}(\bar{n},\bar{T})$ one can show
\begin{eqnarray}
\frac{\dot{\bar{T}}}{\bar{T}}=\left(\frac{\partial \bar{P}}{\partial \bar{\rho}}\right)_n \frac{\dot{\bar{n}}}{\bar{n}}\,, 
\label{tl2}
\end{eqnarray}
giving us the time dependence of temperature. A derivation of the above important relation is given in appendix \ref{appn}.Using the equation of state in Eq.~(\ref{eos}) we can write the above equation as
\begin{eqnarray}
\frac{\dot{\bar{T}}}{\bar{T}}=\omega \frac{\dot{\bar{n}}}{\bar{n}}\,,
\label{tdote}
\end{eqnarray}  
whose solution is \cite{Lima:1995xz}
\begin{eqnarray}
\bar{n}(\bar{T})=n_0 \left(\frac{\bar{T}}{T_0}\right)^{1/\omega}\,.
\label{nt}  
\end{eqnarray}
Using the result from Eq.~(\ref{nre}) in Eq.~(\ref{tdote}) we
immediately get
\begin{eqnarray}
\bar{\rho}(\bar{T})=\rho_0 \left(\frac{\bar{T}}{T_0}\right)^{(1+\omega)/\omega}\,,
\label{rte}
\end{eqnarray}
where in the above equations $n_0$, $T_0$ and $\rho_0$ are the values of the respective variables at some specific time. The above relations show that in presence of radiation fluid when $\omega=1/3$ we have $\bar{\rho}_r \propto \bar{T}^4$ as in standard cosmological models without particle creation.

One can write down the Einstein equations from Eq.~(\ref{efe}) as:
\begin{eqnarray}
  H^2 = \frac{\kappa}{3}\bar{\rho}\,,\,\,\,\,\,
  2\dot{H} + 3H^2= -\kappa (\bar{P} + \bar{P}_c)\,.
\label{gfrw}
\end{eqnarray}
The above two equations can be combined and written as
\begin{eqnarray}
\dot{H}=-\frac32 (1+\omega)H^2\left[1-\frac{{\Gamma}}{3H}\right]\,.
\label{hdote}
\end{eqnarray}
The above equation shows that as long as ${\Gamma} \ll 3H$ one
recovers the standard FLRW solutions in GR. If ${\Gamma} = 3H$,
irrespective of the value of $\omega$ the model produces a de Sitter
phase where $\dot{H}=0$. In this paper we
will focus on the de Sitter phase in presence of radiation. In
general ${\Gamma}$ will be a function of time and the de Sitter
phase can be obtained for a short time interval and after that
interval ${\Gamma} < 3H$ and the system will start to evolve
towards standard radiation dominated cosmological evolution if $\omega=1/3$.
%%%%%%%%%%%%%%%%%%%%%%%%%%%%%%%%%%%%%%%%%%%%%%%%%%%%%%%%%%%%%%%%%%%%%%%%%%%%%%%%%%%%%%%
\section{de Sitter phase in presence of radiation}
\label{dS}

In the present case one can indeed have a de Sitter phase in presence
of radiation as discussed in Ref.~\cite{Lima:2012cm}. In the above
reference the authors briefly mention about such a phase which can
exist in the very early universe. The previous work
  only gave the background cosmological solution, the question related to the 
  metric perturbations on the background solutions was not
  addressed. In the present work we will address the issue related to
  cosmological metric perturbations in de Sitter space in presence of
  radiation. Let us write down the specific equations relevant for
studying this phase when $\omega=1/3$ and the energy density is
$\bar{\rho}_r$ due to radiation.  The equations are
\begin{eqnarray}
  H^2 = \frac{\kappa}{3}\bar{\rho}_r\,,
\label{hrad}
\end{eqnarray}
and
\begin{eqnarray}
\dot{H}=-2 H^2\left[1-\frac{{\Gamma}}{3H}\right]\,,
\label{hdotr}
\end{eqnarray}
where $\Gamma$ is the particle creation rate, in  the de Sitter phase, in the presence of radiation. In this phase we have 
\begin{eqnarray}
\bar{\rho}_r = A \bar{T}^4\,,
\label{rrds}
\end{eqnarray}
where $A=\frac{\pi^2}{30}g_*$, $g_*$ being the effective relativistic internal degrees of freedom.

To proceed further we require a concrete form of the particle
production rate. As discussed in the introduction, there are various
ways in which people have tried to write down a mathematical form of
${\Gamma}$. From our discussion in the last section we have assumed
$\Gamma=\Gamma(H)$ such that it vanishes when expansion of the
universe stops \cite{Lima:2012cm,Pan:2018ibu}i.e., $H=0$ for all
time. In general one can propose that the particle production rate is
given by \cite{Chakraborty:2014fia}
\begin{eqnarray}
{\Gamma}={\Gamma}_0 + lH^2 + mH + \frac{p}{H},\,\,\,\,\,(H>0)
\label{gengam}
\end{eqnarray}
where ${\Gamma}_0,\,l,\,m,\,p$ are constants.  We assumed that
$\bar{\Gamma}_0$ and $\frac{p}{H}$ exists for a non-vanishing Hubble
parameter. Here except $m$ the other constants are dimensional. If
only the first term ${\Gamma}_0$ is present then it will be able to
model an emergent universe scenario where the initial singularity is
avoided.  Cosmological models where ${\Gamma}$ is proportional to
$H^2$ can be applied to the early period of radiation domination. The
universe filled with radiation can proceed towards a de Sitter phase
as we will see in this paper. The other terms in the above general
expression of ${\Gamma}$ like the term proportional to $H$ or the one
proportional to $1/H$ can give the particle production rates in the
early decelerated phase of expansion and later accelerated phase of
expansion, respectively. In general one can combine all of them
together, as done above, to write down the general phenomenological
expression of particle production rate.  In this paper we will follow
the form of ${\Gamma}$ as discussed in Ref.~\cite{Lima:2012cm}. In
this model ${\Gamma}$ is proportional to $H^2$ as we are interested in
the very early universe in the presence of radiation. If we model the
system with particle creation rate as\cite{Lima:2012cm}
\begin{eqnarray}
{\Gamma} = \frac{3 H^2}{H_I}\,,
\label{gamf}
\end{eqnarray}
where $H_I$ is the value of the Hubble parameter when $\dot{H}$ is exactly zero during the initial phase of expansion. In reality $H$ will be near $H_I$ during the de Sitter phase and as long as $H \sim H_I$ the quasi de Sitter phase will exist. As $\dot{H} \sim 0$ in the de Sitter phase $H$ is a slowly varying function and consequently from Eq.~(\ref{hrad}) we see that $\bar{\rho}_r$ is also a slowly varying function during this phase. From Eq.~(\ref{rrds}) it becomes clear that during the quasi de Sitter expansion the temperature of the universe will approximately remain constant. As $\dot{H} \sim 0$ during the de Sitter phase we must have
\begin{eqnarray}
a(t)=a_i e^{H_I t}\,,
\label{sfinf}
\end{eqnarray}
during the quasi de Sitter phase. Here $a_i$ is the scale-factor at
the initial moment of de Sitter phase. To define the quasi de
Sitter phase one must have some variables which mimic the ``slow-roll parameter'' of standard inflationary scenario. In our case we can
define one parameter as
%%%%%%%%%%%%%%%%%%%%%%%%%%%%%%%%%%%%%%%%%%%%%%%%%%%%%%%%%%%%%%%%%%%%
\begin{figure}[t!]
\begin{minipage}[b]{0.5\linewidth}
\centering
\includegraphics[scale=.6]{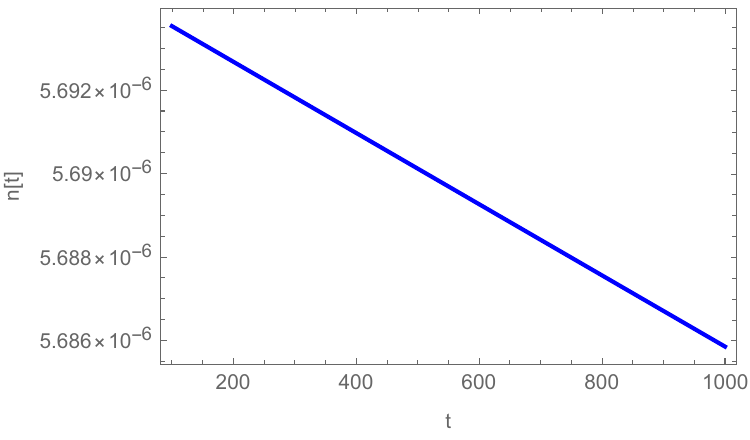}
\caption{Evolution of $n(t)$ with $t$ in the radiation dominated de Sitter phase, more details given in text.} 
\label{n}
\end{minipage}
\hspace{0.2cm}
\begin{minipage}[b]{0.5\linewidth}
\centering
\includegraphics[scale=.6]{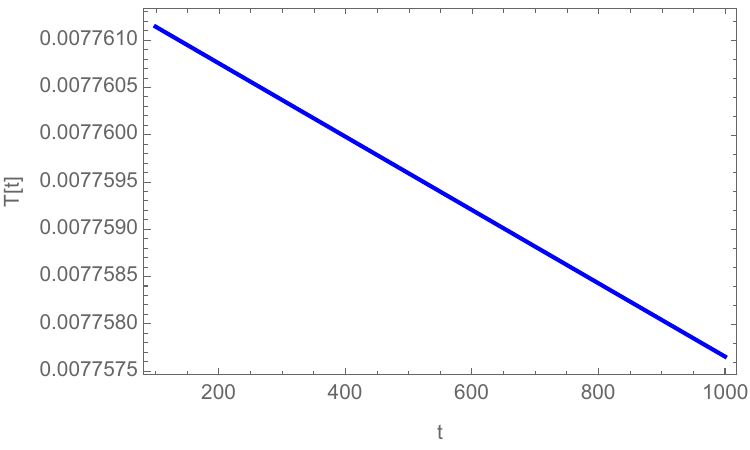}
\caption{Evolution of $T(t)$ with $t$ in the radiation dominated de Sitter phase,  details given text.}
\label{T}
\end{minipage}
\end{figure}
%%%%%%%%%%%%%%%%%%%%%%%%%%%%%%%%%%%%%%%%%%%%%%%%%%%%%%%%%%%%%%%%%%%%%%
\begin{eqnarray}
\epsilon \equiv -\frac{\dot{H}}{H^2}\,, 
\label{srp}
\end{eqnarray}
and assume that quasi de Sitter phase exists as long as
\begin{eqnarray}
\epsilon \ll 1\,.
\label{lsrp}
\end{eqnarray}
From Eq.~(\ref{hdotr}) we can immediately write
\begin{eqnarray}
\epsilon = 2 \left[1-\frac{{\Gamma}}{3H}\right]\,,
\label{ep1}
\end{eqnarray}
which yields
\begin{eqnarray}
H=\left(1-\frac{\epsilon}{2}\right)H_I\,.
\label{hhir}
\end{eqnarray}
In the absence of any potential function for the inflaton field only one slow-roll parameter $\epsilon$ is meaningful. Fig.~\ref{n} and Fig.~\ref{T} show the variation of particle number density and temperature during radiation dominated de Sitter phase in the cosmological model with particle creation where the particle creation rate is given in Eq.~(\ref{gamf}). In producing the plots we have used the initial values of the energy density and number density of particles as
\begin{eqnarray}
\rho_0 = \frac{\pi^2}{30}g_* T_0^4\,,\,\,\,\,n_0=\frac{\zeta(3)}{\pi^2}g_* T_0^4  
\label{intb}
\end{eqnarray}
where $T_0=.008$ and $g_* \sim 100$. Here $\zeta(3)$ is the Riemann-Zeta function of $3$. All the quantities are in Planck units. To get the value of the variables in natural units where all the quantities are expressed in GeV or its inverse powers one has to multiply each quantity with a power of Planck mass where the power is specified by the mass dimension of each quantity. The figures above show that during radiation dominated de Sitter phase the number density and temperature essentially remains constant. This fact is in stern contrast with standard cold inflation scenarios where temperature and number density of particles rapidly vanish to zero. On the other hand these features may be a bit (superficially) similar to warm inflation predictions \cite{Berera:1995ie, Hall:2003zp}.

Before we proceed to the topic of cosmological perturbations in the present model we want to specify an important point regarding the form of the particle production rate ${\Gamma}$. In Eq.~(\ref{gamf}) we proposed phenomenologically a form of ${\Gamma}$ as given in Ref.~\cite{Lima:2012cm} and used it inside the Einstein equation and predicted the background cosmological development. One may invert the process and write down an equivalent form of the particle production rate from the Einstein equation. In this case one does not require to assume any ad hoc form of ${\Gamma}$ to begin with. From Eq.~(\ref{hdotr}) and the definition of the slow-roll parameter $\epsilon$ in Eq.~(\ref{srp}) one can easily write
\begin{eqnarray}
{\Gamma}=3H\left(1-\frac{\epsilon}{2}\right)=\frac{3{\cal H}}{a}\left(1-\frac{\epsilon}{2}\right)\,.
\label{altsrp}  
\end{eqnarray}
Here ${\cal H}$ is the Hubble parameter expressed in conformal time. In this case we have used the Einstein equation and the slow-roll parameter to define the particle production rate. One can easily verify that in the slow-roll approximation
the above form of ${\Gamma}$  is the same as given by the expression in Eq.~(\ref{gamf}). 
%%%%%%%%%%%%%%%%%%%%%%%%%%%%%%%%%%%%%%%%%%%%%%%%%%%%%%%%%%%%%%%%%%%%%%%%%
\section{Scalar cosmological perturbations}
\label{sp}

Following the conventions set in Ref.~\cite{Mukhanov:1990me} we can write the perturbed line element as
\begin{eqnarray}
ds^2 = a^2(\eta)\left\{(1+2\phi)d\eta^2 - 2\partial_i B dx^i d\eta
- \left[(1-2\psi)\delta_{ij} + 2 \partial_i \partial_j E\right]dx^i dx^j \right\}\,,
\label{sclrp}
\end{eqnarray}
where $\phi$, $\psi$, $B$ and $E$ are scalar functions of space and time. Here $d\eta=dt/a$ where $\eta$ is the conformal time. In conformal time
$${\cal H} = \frac{a^\prime}{a}\,,$$
where $a^\prime=da/d\eta$. In this section we will use conformal time exclusively. Henceforth we work in the longitudinal gauge where
$$B=E=0\,.$$ As in the present case the spatial part of the
energy-momentum tensor is diagonal we have $\phi=\psi$. The particle production mechanism particularly  affects the scalar perturbations. To see how it affects various sectors of the theory let us first see how the perturbed continuity equation gets affected in cosmological models with particle production. 

To find out how particle production affects the perturbed continuity equation we calculate
\begin{eqnarray}
\delta(D_\mu {T}^\mu_{{\rm f}\,\nu}) = \partial_\mu \delta {T}^\mu_{{\rm f}\,\nu} + (\delta\Gamma^\mu_{\mu \alpha}) \bar{T}^\alpha_{{\rm f}\,\nu} + \bar{\Gamma}^\mu_{\mu \alpha} \delta{T}^\alpha_{{\rm f}\,\nu}
- (\delta\Gamma^\alpha_{\mu \nu}) \bar{T}^\mu_{{\rm f}\,\alpha} -  \bar{\Gamma}^\alpha_{\mu \nu} \delta{T}^\mu_{{\rm f}\,\alpha}\,.
\end{eqnarray}
Using the expressions of the perturbed affine connections in conformal time we first obtain
\begin{eqnarray}
\delta(D_\mu {T}^\mu_{{\rm f}\,0}) = \delta \rho^\prime + 3{\mathcal H}(\delta \rho + \delta P)
+ (\bar{\rho} + \bar{P})a \nabla^2 V - 3\phi^\prime(\bar{\rho} + \bar{P})\,.
\label{ddtf}
\end{eqnarray}
In our notation
$$\bar{T}^0_{{\rm f}\,0}=\bar{\rho}\,,\,\,\,\,\bar{T}^i_{{\rm f}\,j}=-P \delta^i_j\,,
\bar{T}^i_{{\rm c}\,j}=-P_c \delta^i_j\,,$$
and
$$\delta {T}^0_{{\rm f}\,0}=\delta \rho\,,\,\,\,\,\delta {T}^i_{{\rm f}\,0}=-(\bar{\rho}+\bar{P})a \partial^i V\,,\,\,\,\,\delta {T}^0_{{\rm f}\,i}=-(\bar{\rho}+\bar{P})a^{-1} \partial_i V\,,\,\,\,\,\delta {T}^i_{{\rm f}\,j}=-\delta^i_j \delta P\,,$$
whereas
$$\delta {T}^0_{{\rm c}\,0}=0\,,\,\,\,\,\delta {T}^i_{{\rm c}\,0}=-a
\bar{P}_c\partial^i V\,,\,\,\,\,\delta {T}^0_{{\rm c}\,i}=-\bar{P}_c
a^{-1} \partial_i V\,,\,\,\,\,\delta {T}^i_{{\rm c}\,j}=-\delta^i_j
\delta P_c\,.$$ In the above calculations we use the scalar part
of velocity perturbation given by $\delta u^i =-\partial^i V$ where
$V$ is the fluid velocity potential. Using the above results we can write
\begin{eqnarray}
\delta(D_\mu {T}^\mu_{{\rm c}\,0}) = a \bar{P}_c \nabla^2 V - 3\phi^\prime \bar{P}_c + 3{\mathcal H} \delta P_c\,.
\label{ddtc}
\end{eqnarray}
As $\delta[(D_\mu {T}^\mu_{{\rm f}\,0}) + (D_\mu {T}^\mu_{{\rm c}\,0})]=0$ we have from Eq.~(\ref{ddtc}) and Eq.~(\ref{ddtf})
\begin{eqnarray}
\delta \rho^\prime + 3{\mathcal H}(\delta \rho + \delta P) + (\bar{\rho} + \bar{P})a \nabla^2 V - 3\phi^\prime(\bar{\rho} + \bar{P})+ \bar{P}_c(a \nabla^2 V - 3\phi^\prime)  + 3{\mathcal H} \delta P_c=0\,,
\label{pcon}
\end{eqnarray}
giving us the continuity equation in a cosmological model with
particle creation. One can also find out the Euler equation by working
out $\delta[(D_\mu {T}^\mu_{{\rm f}\,i}) + (D_\mu {T}^\mu_{{\rm
      c}\,i})]=0$. Here $\delta P_c$ is obtained from Eq.~(\ref{crep})
as
\begin{eqnarray}
  \delta P_c = -\frac{a}{3{\mathcal H}}(1+\omega){\Gamma} \,\delta \rho\,.
\label{dpc}
\end{eqnarray}
From the above expression we see that $\delta P_c$ vanishes when ${\Gamma}$ vanishes. When $\Gamma=0$ the above continuity equation reduces to the standard continuity equation in cosmology.
%%%%%%%%%%%%%%%%%%%%%%%%%%%%%%%%%%%%%%%%%%%%%%%%%
\subsection{The other relevant scalar perturbations}

Let us particularly focus on perturbation of $N^\mu$ which gives $\delta N^\mu = \bar{u}^\mu \delta n + \bar{n} \delta u^\mu$ where a bar over the variables specify the unperturbed values.  For the scalar part $\delta u^i =-\partial^i V$ where $V$ is the fluid velocity potential. This gives us
$$\delta N^0 = \bar{u}^0 \delta n\,,\,\,\,\,\,
\delta N^i = -\bar{n} \partial^i V\,.$$
First we perturb the number conservation condition in Eq.~(\ref{pncc}) and get
$$\delta(D_\mu N^\mu)={\Gamma} \delta n\,.$$
Using the fact  $\partial_i \bar{n}=0$ the above equation becomes
$$(\delta n)^\prime + 3{\mathcal H}\delta n +
\bar{n}\delta\Gamma^\mu_{\mu 0} + a\bar{n}\nabla^2 V= a{\Gamma}
\delta n\,,$$ where $\nabla^2 \equiv
-\partial_i \partial^i$. The above equation can also be written as
$$(\delta n)^\prime + (3{\mathcal H} - a{\Gamma})\delta n + \bar{n}\delta\Gamma^\mu_{\mu 0} + a\bar{n}\nabla V= 0\,.$$
We can express Eq.~(\ref{pncc}) in conformal time as
$\bar{n}^\prime + (3{\mathcal H}-a{\Gamma})\bar{n}=0$, which implies
$$(\delta n)^\prime -\frac{\bar{n}^\prime}{\bar{n}}\delta n +
\bar{n}\delta\Gamma^\mu_{\mu 0} + a\bar{n}\nabla^2 V= 0\,.$$ As $\delta \Gamma^\mu_{\mu 0}=-2\phi^\prime$, in the
longitudinal gauge, we can write the final expression of the perturbed
particle number conservation as:
\begin{eqnarray}
(\delta n)^\prime -\frac{\bar{n}^\prime}{\bar{n}}\delta n -2\phi^\prime \bar{n} + a\bar{n}
\nabla^2 V= 0\,.
\label{pdmnm}
\end{eqnarray}
This equation specifies how $\delta n$ evolves with time. This is a coupled partial differential equation and can only be solved if we know the evolution equation for $\phi$ and the form of $\delta \Gamma$.

In the present case the energy momentum tensor is $T^{\mu \nu}=T_{\rm f}^{\mu \nu} + T^{\mu \nu}_{\rm c}$
as given in Eq.~(\ref{efe}) and consequently we have
\begin{eqnarray}
T^0_0= \rho\,,\,\,\,\,\,T^i_j = -(P + P_c)\delta^i_j\,.
\label{tijc}
\end{eqnarray}
Following the method of gauge invariant scalar cosmological perturbations in Ref.\cite{Mukhanov:1990me} one can write the scalar perturbation equations in the longitudinal gauge as:
\begin{eqnarray}
\nabla^2 \phi - 3 {\mathcal H} \phi^\prime -3{\mathcal H}^2 \phi &=& \frac{\kappa}{2} a^2
\delta \rho\,,
\label{scp1}\\
D_i(a\phi)^\prime &=& \frac{\kappa}{2}(\bar{\rho} + \bar{P} + \bar{P}_c) a^2 \delta u_i\,,
\label{scp2}\\
\phi^{\prime \prime} + 3 {\mathcal H} \phi^\prime + (2{\mathcal H}^\prime + {\mathcal H}^2)\phi
&=& \frac{\kappa}{2} a^2 (\delta P + \delta P_c)\,.
\label{scp3}
\end{eqnarray}
The mathematical form of $\delta P_c$ is given in Eq.~(\ref{dpc}). Using the
fact that $\delta P = \omega \delta \rho$ one can simplify the above
equations as
$$\phi^{\prime \prime} + 3 {\mathcal H} \phi^\prime + (2{\mathcal
  H}^\prime + {\mathcal H}^2)\phi=\frac{\kappa}{2} a^2 \omega \delta
\rho + \frac{\kappa}{2} a^2 \delta P_c\,.$$ For the sake of generality
we will try to keep the general equation of state explicit in the
equations written below. Only when we state our results we will use
$\omega=1/3$ as we are working primarily in a universe filled with
radiation.  One can use the form of $(\kappa/2)a^2\delta \rho$ from
Eq.~(\ref{scp1}) in the first term on the right hand side of the above
equation and obtain:
\begin{eqnarray}
\phi^{\prime \prime} - \omega \nabla^2 \phi + 3{\mathcal H}(1+\omega)\phi^\prime
+[2{\mathcal H}^\prime + {\mathcal H}^2(1+3\omega)]\phi = \frac{\kappa}{2} a^2 \delta P_c\,.
\label{scl1}
\end{eqnarray}
Using the expression of $\delta P_c$ one can write the above expression as
\begin{eqnarray}
\phi^{\prime \prime} - \left[\omega -\frac{a(1+\omega){\Gamma}}{3{\mathcal H}}\right]\nabla^2 \phi
&+& \left(3{\mathcal H}-a{\Gamma}\right)(1+\omega)\phi^\prime
+[2{\mathcal H}^\prime + {\mathcal H}^2(1+3\omega)\nonumber\\
&-&a(1+\omega){\Gamma}{\mathcal H}]\phi = 0\,.
\label{finsp}
\end{eqnarray}
This equation specifies the growth of $\phi$ in presence of gravitational particle production. From Eq.~(\ref{scp2}) we can write
\begin{eqnarray}
\nabla^2 V = -\frac{2}{\kappa a^2}\frac{a^\prime \nabla^2 \phi + a \nabla^2 \phi^\prime}{(\bar{\rho} + \bar{P} + \bar{P}_c)}\,.
\label{Veqn1}
\end{eqnarray}
Using this equation we can write Eq.~(\ref{pdmnm})
as
\begin{eqnarray}
\delta n^\prime -\frac{\bar{n}^\prime}{\bar{n}}\delta n -\left[2\phi^\prime  +
    \frac{2 ({\mathcal H} \nabla^2 \phi + \nabla^2\phi^\prime)}{\kappa(\bar{\rho} + \bar{P} + \bar{P}_c)}\right]\bar{n} = 0\,.
\label{pdmnm1}
\end{eqnarray}  
These above equations specify the scalar perturbation equations of the cosmological model with particle creation. These equations can be applied to any cosmological model where gravitational particle creation takes place. 
%%%%%%%%%%%%%%%%%%%%%%%%%%%%%%%%%%%%%%%%%%%%%%%%%%%%%%%%%%%%%%%%%%
\subsection{Evolution of perturbations}

Before we write down the evolution equation of the perturbations using the above fact we want to opine on the temperature growth of the background and some properties of temperature fluctuations.  Using Eq.~(\ref{tdote}) and the expression of $\dot{n}/n$ we can write
\begin{eqnarray}
\frac{\bar{T}^\prime}{\bar{T}} =\omega \frac{\bar{n}^\prime}{\bar{n}}
=\omega(a{\Gamma} -3 {\mathcal H})\,,
\label{tpbt}
\end{eqnarray}
where primes naturally denote differentiation with respect to conformal time $\eta$. The temperature fluctuations can be approximately figured out by assuming that the metric perturbations do not take the system out of local thermal equilibrium. Assuming the perturbed system to be  near local thermodynamic equilibrium one can still assume 
\begin{eqnarray}
\rho = \bar{\rho} + \delta \rho  \sim \frac{\pi^2}{30}g_* T^4
\label{rtr}
\end{eqnarray}
where $T=\bar{T} + \delta T$. Up to first order the relation gives
\begin{eqnarray}
\delta T \sim \frac{15}{2\pi^2 g_*}\frac{\delta \rho}{\bar{T}^3}\,.
\label{dte}
\end{eqnarray}
%%%%%%%%%%%%%%%%%%%%%%%%%%%%%%%%%%%%%%%%%%%%%%%%%%%%%%%%%%%%%%%%%%%%
\begin{figure}[t!]
\begin{minipage}[b]{0.5\linewidth}
\centering
\includegraphics[scale=.6]{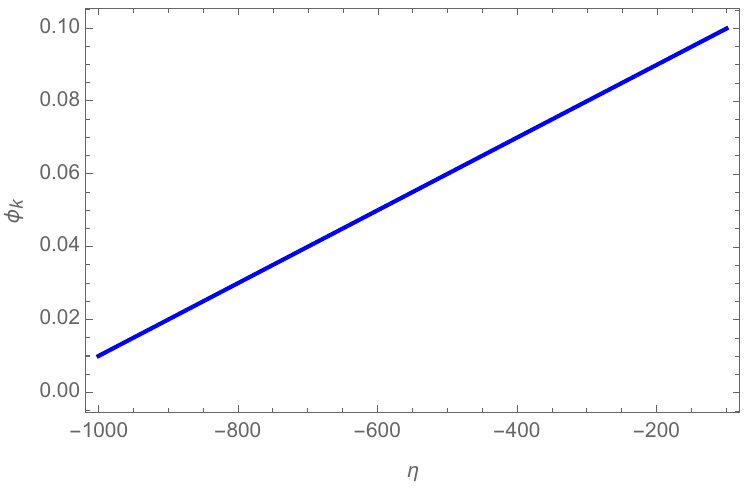}
\caption{Evolution of $\phi_k$ with $\eta$ in the radiation dominated de Sitter phase for $\delta \Gamma=0$ and $k=10^{-5}$, more details given in text.} 
\label{phi}
\end{minipage}
\hspace{0.2cm}
\begin{minipage}[b]{0.5\linewidth}
\centering
\includegraphics[scale=.6]{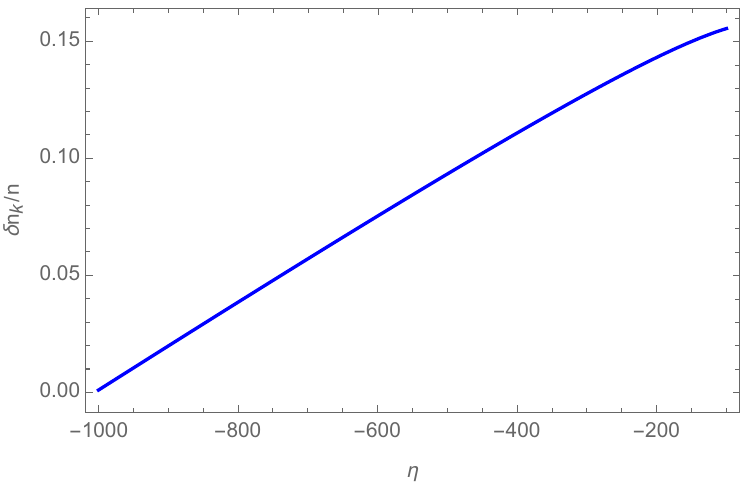}
\caption{Evolution of $\delta n_k/n$ with $\eta$ in the radiation dominated de Sitter phase for $\delta \Gamma=0$ and $k=10^{-5}$,  details given text.}
\label{dnn}
\end{minipage}
\end{figure}
%%%%%%%%%%%%%%%%%%%%%%%%%%%%%%%%%%%%%%%%%%%%%%%%%%%%%%%%%%%%%%%%%%%%%%
We will see later that metric perturbations may induce fluid flow in the initial stages of the de Sitter phase and consequently local thermodynamic equilibrium can only hold approximately. The above simple estimation can give us the temperature fluctuations induced by metric perturbations assuming the system is near about local thermal equilibrium.

In this paper we have assumed that radiation fluid is the main constituent of the universe, where $\omega=1/3$, so that Eq.~(\ref{tpbt}) can easily be integrated to see how the background temperature evolves when one has
\begin{eqnarray}
{\Gamma}=\frac{3{\mathcal H}^2}{a^2 H_I}\,,
\label{gaminc}
\end{eqnarray}
which can be obtained from Eq.~(\ref{gamf}).  For plane wave like perturbations where the perturbation modes are proportional to $e^{i{\bf k}\cdot{\bf x}}$ we will have the basic perturbation equations as:
\begin{eqnarray}
\phi_k^{\prime \prime} + k^2\left[\omega -\frac{a(1+\omega){\Gamma}}{3{\mathcal H}}\right]\phi_k
&+& (3{\mathcal H}-a{\Gamma})(1+\omega)\phi_k^\prime
+[2{\mathcal H}^\prime + {\mathcal H}^2(1+3\omega)\nonumber\\
&-&a(1+\omega){\Gamma}{\mathcal H}]\phi_k = 0\,,
\label{finsp2}
\end{eqnarray}
%%%%%%%%%%%%%%%%
\begin{eqnarray}
\delta n_k^\prime -\frac{\bar{n}^\prime}{\bar{n}}\delta n_k -2\left[\phi^\prime_k  -
    \frac{k^2({\mathcal H} \phi_k + \phi^\prime_k)}{\kappa(\bar{\rho} + \bar{P} + \bar{P}_c)}\right] \bar{n} &=& 0\,,
\label{pdmnm2k}\\
V_k + \frac{2}{\kappa a^2}\frac{a^\prime \phi_k + a \phi^\prime_k}{(\bar{\rho} + \bar{P} + \bar{P}_c)} &=& 0\,,
\label{Veqn2k}
\end{eqnarray}
where $\phi_k(\eta)$, $\delta n_k(\eta)$ and $V_k(\eta)$ are the $k$th Fourier modes of the respective perturbations. Using the following expression
\begin{eqnarray}
\bar{\rho} + \bar{P} + \bar{P}_c =(1+\omega)\bar{\rho}\left[1-\frac{a {\Gamma}}{3{\mathcal H}}\right]=\frac{3{\mathcal H}^2}{\kappa a^2}(1+\omega)\left[1- \frac{\mathcal H}{a H_I}\right]\,,
\nonumber
\end{eqnarray}
we can write Eq.~(\ref{pdmnm2k}) and Eq.~(\ref{Veqn2k}) as
\begin{eqnarray}
\delta n_k^\prime -\frac{\bar{n}^\prime}{\bar{n}}\delta n_k -2\left[\phi^\prime_k  -
    \frac{a^2 k^2({\mathcal H} \phi_k + \phi^\prime_k)}{3 {\mathcal H}^2(1+\omega)(1 - \frac{\mathcal H}{a H_I})}\right]\bar{n} &=& 0\,,
\label{pdmnm3k}\\
V_k + \frac{2}{3 {\mathcal H}^2}\frac{a^\prime \phi_k +  a \phi^\prime_k}{(1+\omega)(1- \frac{\mathcal H}{a H_I})} &=& 0\,.
\label{Veqn3k}
\end{eqnarray}
In the present case one can write ${\mathcal H}^\prime$ and ${\mathcal H}$ as functions of the slow-roll parameter as
\begin{eqnarray}
{\mathcal H}^\prime = (1- \epsilon){\mathcal H}^2\,,\,\,\,\,
\frac{\mathcal H}{a H_I}=1-\frac{\epsilon}{2}\,.  
\label{ep1c}
\end{eqnarray}
%%%%%%%%%%%%%%%%%%%%%%%%%%%%%%%%%%%%%%%%%%%%%%%%%%%%%%%%%%%%%%%%%%%%
\begin{figure}[t!]
\centering
\includegraphics[scale=.6]{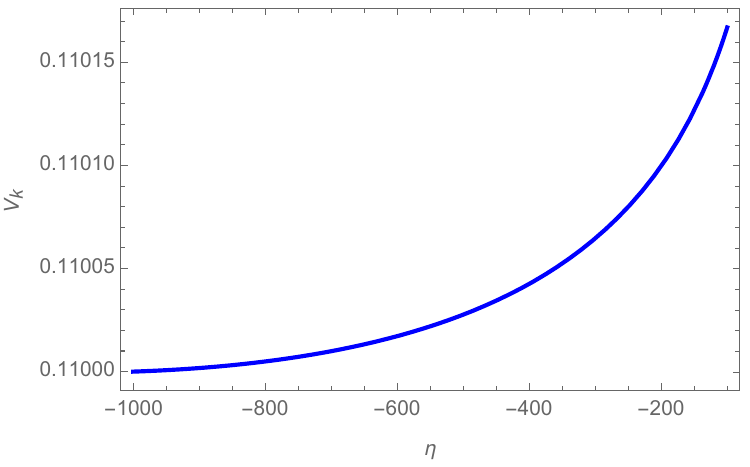}
\caption{Evolution of $V_k$ with $\eta$ in the radiation dominated de Sitter phase for $\delta \Gamma=0$ and $k=10^{-5}$.} 
\label{v}
\end{figure}
%%%%%%%%%%%%%%%%%%%%%%%%%%%%%%%%%%%%%%%%%%%%%%%%%%%%%%%%%%%%%%%%%%%%%%
The differential equations show the nature of the scalar perturbation for $k  < {\mathcal H}$, which corresponds to the superhorizon limit and  $k  > {\mathcal H}$, corresponding to the subhorizon limit. The perturbation equations can be compactly written down using the slow-roll parameter as:
\begin{eqnarray}
\phi_k^{\prime \prime} + \frac32(1+\omega){\mathcal H} \epsilon\phi_k^\prime &+& \left[\frac{\epsilon}{2}k^2(1+\omega)-k^2 
  + 2{\mathcal H}^\prime + {\mathcal H}^2(1+3\omega)\right.\nonumber\\
& &\left. -3(1+\omega){\mathcal H}^2(1-\frac{\epsilon}{2})\right]\phi_k = 0\,,
\label{finsp3}
\end{eqnarray}
and
\begin{eqnarray}
\delta n_k^\prime +\frac32 \epsilon {\mathcal H} \delta n_k - 2\left[\phi^\prime_k  -
    \frac{2 a^2 k^2({\mathcal H} \phi_k + \phi^\prime_k)}{3 {\mathcal H}^2(1+\omega)\epsilon} \right]\bar{n} &=& 0\,,
\label{pdmnm4k}\\
V_k + \frac{4}{3 {\mathcal H}^2}\frac{a^\prime \phi_k +  a \phi^\prime_k}{(1+\omega)\epsilon} &=& 0\,.
\label{Veqn4k}
\end{eqnarray}
%%%%%%%%%%%%%%%%
It must be noted that $\delta n_k \ll \bar{n}$ but in an absolute sense $\delta n_k$ need not be smaller than one as the fluctuation in the number of particles in unit volume cannot be less than one. The above gravitational and hydrodynamic perturbation equations are distinctly different from standard perturbation equations in radiation domination or matter domination or from the perturbation equations in warm inflation or super cold inflation. In standard perturbation equations in presence of matter the concept of slow-roll never arises and in inflationary models the concept of slow-roll arises from the nature of inflaton potential. In the present case we have a slow-roll mechanism in presence of hydrodynamic fluid where the slowness of the process is related to the slow decrement of the particle creation rate which sustained a near de Sitter phase initially.   

The plots of the perturbations are shown in Fig.~\ref{phi},
Fig.~\ref{dnn} and Fig.~\ref{v}. In all of these plots we have taken
$k=10^{-5}$ and $\omega=1/3$. We have used Planck units as described earlier. While
plotting the above figures we used $T_0=.008$ and $g_* \sim 100$ and
${\mathcal H}_0=.001$. The zero subscript specify initial value of any
variable. The scale-factor initially is taken to be unity. For the
perturbations we have used $\phi_k(\eta_0)=.01$,
$\phi_k^\prime(\eta_0)=.0001$, and $\delta n_k(\eta_0)=.001\times n_0$ where
$n_0$ is specified in Eq.~(\ref{intb}). In all of the plots
$\eta_0=-10^3$.  The plots show well behaved perturbations of the
relevant quantities in the radiation dominated de Sitter phase.
%%%%%%%%%%%%%%%%%%%%%%%%%%%%%%%%%%%%%%%%%%%%%%%%%%%%%%%%%%%%%%%%%%%%
\begin{figure}[t!]
\centering
\includegraphics[scale=.6]{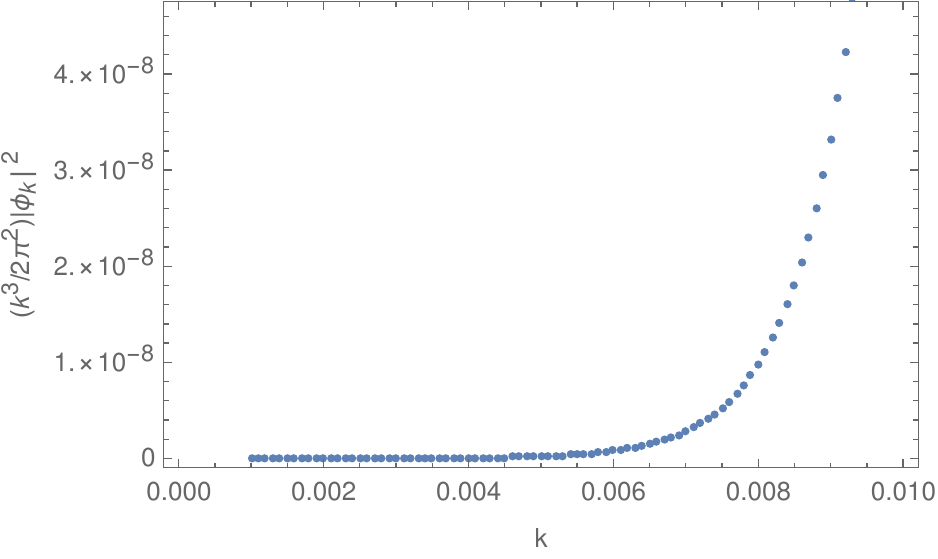}
\caption{Plot of $\frac{k^3}{2\pi^2}|\phi_k|^2$ with comoving wave number $k$. The plot does not indicate a nearly scale invariant power spectrum as expected from simple inflationary models.} 
\label{power}
\end{figure}
%%%%%%%%%%%%%%%%%%%%%%%%%%%%%%%%%%%%%%%%%%%%%%%%%%%%%%%%%%%%%%%%%%%%%%

The above perturbation equations may produce unstable behavior when $k>{\mathcal H}$. The perturbations do not turn out to be singular in this regime but tend to be unstable and may slowly attain value near to or greater than one. From Eq.~(\ref{finsp3}) it is seen that when $k^2>{\mathcal H}^2$, $\phi_k$ can become  unstable\footnote{In this limit we must also have $k^2 > {\mathcal H}^\prime$ as ${\mathcal H}^\prime = (1- \epsilon){\mathcal H}^2$ and $\epsilon \ll 1$.}. To see this explicitly we write Eq.~(\ref{finsp3}) when $\epsilon \ll 1$ as
\begin{eqnarray}
\phi_k^{\prime \prime} \sim k^2 \phi_k\,,\,\,\,\,{\rm when}\,\,\,k>{\mathcal H}
\label{phiku}
\end{eqnarray}
showing that $\phi_k$ can increase without bound in this limit. The
condition $k^2>{\mathcal H}^2$ will not stay forever and consequently
$\phi_k$ will not become singular. It is seen that $\phi_k$ can indeed
become non-perturbative, particularly if the initial value of it is
not too small, when $k^2>{\mathcal H}^2$. On the other hand when  $k^2 < {\mathcal H}^2$
and $\epsilon \ll 1$, Eq.~(\ref{finsp3}) becomes $\phi_k^{\prime
  \prime} \sim 2( {\mathcal H}^2-{\mathcal H}^\prime )\phi_k =
\epsilon {\mathcal H}^2 \phi_k$, and in our approximation scheme we
can write the last condition as
\begin{eqnarray}
\phi_k^{\prime \prime} \sim 0\,,\,\,\,\,{\rm when}\,\,\,k<{\mathcal H}
\label{phikd}
\end{eqnarray}
showing that $\phi_k$ will be approximately a linear function of conformal time in this limit. In particular if $\phi_k^\prime \sim 0$ then the scalar perturbation will be a constant in this limit. As $\epsilon \to 0$ when $\bar{\Gamma}\to 3{\mathcal H}/a$ we see Eq.~(\ref{pdmnm4k}) can also be written as
\begin{eqnarray}
\left(\frac{\delta n_k}{\bar{n}}\right)^\prime  \sim  2\left[\phi^\prime_k  -
    \frac{2 a^2 k^2({\mathcal H} \phi_k + \phi^\prime_k)}{3 {\mathcal H}^2(1+\omega)\epsilon} \right]\,.
\label{nks}
\end{eqnarray}
From the above expression we see the fractional number density can build up if $\phi_k$ increases in the limit $k>{\mathcal H}$. Depending upon the sign of $\phi^\prime_k$ and $\phi_k$ we see that $\delta n_k/\bar{n}$ can be positive or negative implying that at some length scales there will be more fractional particle and at some length scales the particle fraction is less. These results show that one can have inhomogeneities in the radiation filled universe during the de Sitter phase. 

Before we end the discussion on scalar cosmological perturbations in de Sitter space in presence of radiation we want to comment on the nature of the gravitational potential power spectrum. We know that the simplest inflationary models predicts nearly scale invariant power spectrum. In the present case we have seen that perturbation with high $k$ tend to become nonlinear and we expect the power spectrum to be not nearly scale invariant. In the present case we plot $(k^3/2\pi^2)|\phi_k|^2$ with $k$ in Fig.~\ref{power}. In Fig.~\ref{power} we have assumed the slow-roll parameter $\epsilon=.001$ and have plotted the perturbations calculated at $\eta=-500$. The initial Hubble parameter used to plot the perturbations is ${\cal H}=.0001$. The initial values of the perturbation and its time derivative remain the same as stated before. Both $\eta$ and Hubble parameter are expressed in Planckian units. We see that the nature of the curve does not portray a nearly scale invariant power spectrum. In the high $k$ region the curve is approximately proportional to $k^3$. From this figure it is apparent that the present model of de Sitter space in presence of radiation, accompanied by  gravitational particle production,  does not reproduce the simplest features seen in the simple inflationary models.   
%%%%%%%%%%%%%%%%%%%%%%%%%%%%%%%%%%%%%%%%%%%%%%%%%%%%%%%%%%%%%%%%%%%%%%%%%%%%%%%%%
\section{Brief comment on the other kind of metric perturbations}
\label{vt}

The de Sitter phase in a radiation filled universe can produce
interesting features in pure vector perturbation sector. The tensor
perturbations on the other hand do not produce any novel features. The
pure vector fluid velocity perturbation can get amplified due to the
particle production process inducing radiation fluid movement in the
shorter length scales whereas tensor perturbations in the present
case do not get amplified and its properties remain similar to the
corresponding result in standard super cool inflationary scenario. In
this section we will discuss briefly about the vector and tensor
perturbations in cosmology with particle creation. The basic results
related to the metric perturbations quoted in the present section
follows the conventions set in Ref.~\cite{Mukhanov:1990me}.
%%%%%%%%%%%%%%%%%%%%%%%%%%
\subsection{Vector perturbations}

When only vector perturbations are excited the metric is given as
\[
\renewcommand\arraystretch{1.3}
 g_{\mu\nu}=a^2(\eta)
\left[
\begin{array}{c|c}
  1 & S_{i} \\
  \hline
  S_{i} &  -\delta_{ij}+\partial_jF_i+\partial_iF_j
\end{array}
\right]\,,
\]
where $S^i$ and $F^i$ are 3-vectors satisfying the constraint
$\partial_iS^i = \partial_i F^i=0$. In the Newtonian gauge where
$F^i=0$, the standard theory of vector perturbation gives
\begin{equation}
  v^i_k = \frac{k^2}{2a^2  \kappa (\bar{\rho} + \bar{P} + \bar{P}_c)} S^i_k
  = \frac{k^2}{a^2  \kappa (1+\omega)\bar{\rho}\epsilon} S^i_k
  = \frac{k^2}{3{\mathcal H}^2 (1+\omega) \epsilon} S^i_k\,,
\label{vik}  
\end{equation}
and
\begin{equation}
S^i_k(\eta) = \frac{C^i_k}{a(\eta)^2}\,,
\label{sik}  
\end{equation}
where $C^i_k$ is a constant 3-vector. Here
$v^i_k$ is the ${\bf k}$th Fourier component of the pure vector part of the
fluid velocity 3-vector $v^i$. Here $S^i_k$ designates the
${\bf k}$th Fourier mode of the metric vector perturbation. For all these
variables ${\bf k}\cdot {\bf v}_k={\bf k}\cdot {\bf S}_k = {\bf k}\cdot {\bf C}_k=0\,.$
In this case we see that the metric perturbation $S^i_k$ becomes negligible as the universe expands, but from Eq.~(\ref{vik}) it is seen that the initial values of $v^i_k$ may not be negligible when $\epsilon$ is small and $k > {\mathcal H}$. The pure vector part of fluid velocity perturbations cannot be neglected in this case and this may have important cosmological consequences. In pure radiation domination the pure vector part of fluid velocity perturbations remains constant, and if they were negligible initially they remain so throughout the radiation dominated phase. In the present case the pure vector part of fluid velocity perturbations cannot be neglected in the de Sitter phase.
%%%%%%%%%%%%%%%%%%%%%%%%%%%%%%%%%%%%
\subsection{Tensor perturbations}

In order to tackle tensor perturbations
\begin{eqnarray}
ds^2 = a^2(\eta)\left[d\eta^2 - (\delta_{ij} - h_{ij})dx^i dx^j\right]\,,
\label{gten}
\end{eqnarray}
where $h_{ij}$ is a symmetric, traceless tensor with $\partial^j h_{ij}=0$ one can rescale $h_{ij}$ and write 
\begin{eqnarray}
h_{ij}=\frac{v}{a}e_{ij}\,,   
\label{hije}
\end{eqnarray}
where $v$ is a function of $\eta\,,\,{\bf x}$ and $e_{ij}$ is the time-independent polarization tensor. Expanding $v$ in plane waves we get,
\begin{eqnarray}
v_k^{\prime\prime} + \left(k^2 -\frac{a^{\prime\prime}}{a}\right)v_k=0\,.
\label{vke}
\end{eqnarray}
In our case during the initial stages of quasi-de Sitter expansion $v_k$ has a solution which is similar to the solution of the analogous problem in de Sitter space in the absence of any radiation. The tensor perturbations do not carry any special signature from the particle production model up to linear order of perturbations. For non-linear orders the modified scalar perturbations can act as source for the tensor modes and consequently, in general, the tensor power spectrum will also differ from the standard inflationary result.  In the present case once the particle production rate decreases considerably, normal radiation domination returns when $a(\eta) \propto \eta$  and then
$v_k \propto \exp(\pm ik\eta)$.

The presence of the source term ${\Gamma}$, where
  the form of ${\Gamma}$ as given in Eq.~(\ref{gamf}), in the
  perturbation equations modify the behavior of the evolution of
  cosmological perturbations from pure radiation dominated universe.
  Cosmological particle production in the slow-roll regime introduces
  finite instabilities in the short wavelength limit of the
  perturbations. The nature of the instabilities greatly depend upon
  the form of ${\Gamma}$ chosen for the background particle
  production rate. The physical origin of the instabilities can be
  intuitively understood. In de Sitter space, filled with radiation
  fluid, one has a casual Hubble patch. The radius of the Hubble
  sphere defines the superhorizon and subhorizon limits. Events inside
  the Hubble patch are understood to be casually interrelated. In
  general it is very difficult for a hydrodynamic fluid to exist for
  some time in a de Sitter space as due to exponential increase of the
  scale-factor, the fluid density will rapidly fade. In our model the
  radiation bath persists as there is a constant rate of gravitational
  particle production per unit spatial volume. In absence of fluid
  velocity fluctuations there is no instability of the model and as a
  result the background model is perfectly stable. As soon metric
  perturbations are switched on there appears (scalar) fluid velocity
  fluctuations which can deplete some regions of the universe and
  overpopulate some causally connected region of the universe. These
  overpopulated regions have larger gravitational effect and attract
  more fluid from the neighborhood, where particle creation is
  happening in an unperturbed manner, and as a result we have unstable
  perturbations in the causally connected subhorizon limit. Even the
  pure vector velocity perturbation can become appreciably big inside
  the subhorizon region. In our model the particle production rate is
  not perturbed, remains practically constant during the slow-roll
  phase, and as a result some regions can continuously supply
  particles to other dense regions. The instabilities persist in the
  subhorizon scale because depletion and overpopulation are causally
  connected and primarily affects the system in the subhorizon scales. This
  intuitive interpretation of the model show that the instabilities we
  see in our model are a natural consequence of the assumptions we have made to define our cosmological model.
%%%%%%%%%%%%%%%%%%%%%%%%%%%%%%%%%%%%%%%%%%%%%%%%%%%%%%%%%%%%%%%%%%%%%%%%%%%%%%%%%%
\section{Discussion and Conclusion}

In this article we have tried to formulate metric perturbations in a cosmological model where gravitational particle production is taking place. The formulation requires various new assumptions about the particle production process in cosmology. Previous authors have presented the cosmological background (unperturbed) calculations of radiation filled universe in a de Sitter phase where the de Sitter negative pressure is supplied by the particle creation pressure. If the particle creation rate attains a specific value then a radiation filled universe may enter a de Sitter phase where one may try to see whether such a phase can be modelled by a slow-roll process. The expression of the creation pressure turns out to be proportional to the particle production rate and energy density and always remains negative. We have used the expression connecting the creation pressure and energy density like a new equation of state from which one can get the perturbed creation pressure. In the initial part of the present paper we have presented the basic equations governing the background cosmological evolution in a cosmological model with particle production. It has been pointed out how such a cosmological model can enter a de Sitter phase in presence of radiation. In the model of particle production, as studied in the paper, specific entropy does not increase and the created radiation can locally equilibrate with the existing radiation, although the whole system is undergoing an irreversible process where total entropy increases due to the enlargement of the phase space. As long as the radiation filled universe remains quasi de Sitter, massless particle production can go on, and as slowly the system comes out of the de Sitter phase massless particle production goes down as predicted by Leonard Parker \cite{Parker:1969au}. Following previous works on the cosmology of de Sitter space in presence of radiation we have assumed the particle production rate is proportional to the square of the Hubble parameter.   

The main results of the paper are related to the perturbation of the cosmological model with gravitational particle creation. Initially the scalar cosmological perturbation calculations are presented. We have consistently used the longitudinal gauge in our calculations for the scalar perturbations. While working out the scalar perturbations we have set the perturbation of the particle production rate to be zero. This assumption is based on the fact that in the gravitational particle production models the particle production rate is primarily a function of the expansion rate of the universe which is given by the Hubble parameter. As the expansion rate is not affected by the metric perturbations it is natural to assume that the perturbation of the particle production rate due to metric fluctuations is zero. With this assumption and the equation of state connecting creation pressure and energy density one can write down the scalar perturbation equations in a cosmological model accompanied by gravitational particle production. When the equations are solved for the de Sitter universe, filled with radiation, various interesting results come out. The perturbation equations show that all the relevant Fourier modes of the perturbations in longitudinal gauge show  instability in the subhorizon limit where as on the superhorizon case the Fourier mode of the metric perturbation in the longitudinal gauge linearly increases with time or becomes constant. In general for a radiation dominated universe the short wavelength perturbations oscillate in a stable fashion. The effect of the de Sitter phase is to make the small scale perturbations unstable which may cause inhomogeneities in the very early universe. We have shown that the scalar perturbation  power spectrum steeply rises with the comoving wave number which predicts there is more power for high values of $k$. The exact value of $k$ after which the steepness increases cannot be ascertained as this value depends on the initial condition of the perturbations set on an arbitrary time slice. This feature distinguishes our model from simple inflationary models. 

In the previous section we have briefly discussed the fate of pure vector and tensor perturbations in cosmology where gravitational particle production takes place. The vector metric perturbation dies down quickly in an exponentially expanding spacetime where as the pure vector velocity perturbations may not die down so easily. It has been shown that in the de Sitter phase the fluid pure vector perturbation may be amplified in the subhorizon limit indicating directed flow of radiation from one region to other. These strong local flows can produce interesting effects in the de Sitter model studied. Further investigation is required to unravel the effects of such flow in the universe pervaded by radiation fluid. The tensor part of the perturbation sector does not carry any special signature of gravitational particle production process, there are no amplification or instability in this sector. The tensor perturbations in the present model are similar to the tensor perturbations in standard inflationary theory.

A serious point concerning the present model of cosmology is related to the ultraviolet instability of the scalar cosmological perturbations. The small scale or high $k$ modes of the scalar perturbations become perturbatively unstable. This may mean production of primordial black holes or scalar induced gravitational waves \cite{Domenech:2021ztg}. As the present theory does not require any specific inflaton like scalar field to produce the de Sitter phase so one naturally does not explicitly think about the quantum nature of the scalar fluctuations. The whole theory can be thought of as a classical theory. The difficulty of this classical interpretation is related to the cause of the seed fluctuations. In standard inflationary theory the seed fluctuations are inherent quantum fluctuations which are unavoidable in any realistic physical system. In the present case one faces considerable difficulty if the scalar cosmological perturbations are assumed to be quantized. The quantum nature of the perturbations can naturally answer the question about origin of the seed perturbations but the quantization scheme becomes obscure in the present model. Any quantization scheme of scalar fields will require a vacuum to be defined. In standard inflationary theories the vacuum is called the Bunch-Davies vacuum. This vacuum is defined in such a way that on the ultraviolet (sub-Hubble) scales the perturbation modes become plane-wave modes. Any similar method of choosing a vacuum for the scalar perturbations in the present model faces a problem because in the ultraviolet limit the theory does not remain perturbative.

One possible way to evade the above problem is to assume a pre-de Sitter phase where one can quantize the perturbations in a manageable way. In the present model it is seen that if one keeps the particle production rate  as given in Eq.~(\ref{gamf}) then at  times when $H>H_I$ (assuming an expanding universe where $H>0$) one obtains the phantom phase in presence of radiation. This fact  can be verified from Eq.~(\ref{hdotr}) which shows that $\dot{H}>0$ for $H>H_I$. Whether the perturbation modes in that kind of an exotic phantom phase remains perturbative, in the ultraviolet limit, is an open question. It is difficult to guess an answer here because there are no works related to quantum fluctuations in a phantom phase in presence of radiation. In Ref.~\cite{Carroll:2003st} the authors
have presented a field theoretic interpretation of cosmological perturbations in the phantom phase in presence of a scalar field. In this reference it is
shown that the perturbation theory of the scalar field, which is responsible for the phantom phase, requires an ultraviolet cutoff. In our case the phantom phase is not produced by any scalar field and consequently one has to formulate the perturbation theory in a new context. Here we do not present any details about this reformulation as it is out of scope of the present paper. The present paper shows that the construction of a cosmological model based on a de Sitter phase in presence of radiation is incomplete as the question related to the initial conditions of the perturbations remains unanswered. A focussed research in this field in the future can resolve this issue.

In conclusion we can say that the present work shows that although the particle production model in de Sitter phase can produce all the kinds of metric perturbations produced in inflationary models but it cannot reproduce all the features of the simplest inflationary paradigm. One has to modify the present model to make the scalar power spectrum nearly scale invariant and make the theory quantum mechanically complete.
%%%%%%%%%%%%%%%%%%%%%%%%%%%%%%%%%%%%%%%%%%%%%%%%%%%%%%%%%%%%%%%%%%%%%%%%%%%
\appendix
\section{Appendix I}
\label{appn1}
%%%%%%%%%%%%%
In an expanding universe the energy density naturally decreases with time. In presence of particle production the decrement in energy density, with time, slows down (compared to the corresponding rate in standard cosmological models) due to the production of new particles which contributes to the energy density. This decrement in the rate of energy density, in models with particle production, can be explained if we take into account the action of the creation pressure $P_c$. To understand this point we must note that the effective pressure of the system is the sum of the fluid pressure and the creation pressure,
$$P_T=P+P_c\,,$$
where $P=\omega \rho$ is the fluid pressure. The total pressure appears in the total energy momentum tensor appearing in Eq.~(\ref{efe}). As total energy momentum tensor is conserved, we must have
$$\frac{d(\rho \Delta V)}{dt}=-P_T\frac{d(\Delta V)}{dt}\,.$$
Assuming $d(\Delta V)/dt>0$ for an expanding universe, we can decrease the time rate of change of the energy (compared to standard cosmological models where there is no particle production and $P_T=P$) by making  $P_T<P$. For positive values of $P_T$ and $P$ one can get $P_T<P$ when the creation pressure $P_c$ is negative. As because by introducing the negative creation pressure we can decrease the rate of decrement of the of energy (compared to the case where no particles are created) we do not again require to alter the energy density. This is the reason why we do not alter the energy density for the cosmological scenarios where there is particle production.
%%%%%%%%%%%%%%%%%%%%%%%%%%%%%%%%%%%%%%%%%%%%%%%%%%%%%%%%%%%%%%%%%%%%%%%%
\section{Appendix II}
\label{appn}
%%%%%%%%%%%%
Assuming $\bar{T},\bar{n}$ to be the basic thermodynamic variables we have
$\bar{\rho}=\bar{\rho}(\bar{n},\bar{T})$. As a consequence
$$\dot{\bar{\rho}}=\left(\frac{\partial \bar{\rho}}{\partial
  \bar{n}}\right)_T \dot{\bar{n}} + \left(\frac{\partial
  \bar{\rho}}{\partial \bar{T}}\right)_n \dot{\bar{T}}\,.$$ Using the
above relation in
$\dot{\bar{\rho}}+3H(\bar{\rho}+\bar{P})={\Gamma}(\bar{\rho}+\bar{P})$,
which comes from Eq.~(\ref{nec}), one gets
$$\left(\frac{\partial \bar{\rho}}{\partial \bar{n}}\right)_T \dot{\bar{n}}
+ \left(\frac{\partial \bar{\rho}}{\partial \bar{T}}\right)_n \dot{\bar{T}}=-3H(\bar{\rho} +
\bar{P}) - 3H\bar{P}_c\,.$$
Using the relation of $\dot{\bar{n}}/\bar{n}=\Gamma-3H$, the above equation can also be written as
$$\left(\frac{\partial \bar{\rho}}{\partial \bar{T}}\right)_n \dot{\bar{T}}=-3H(\bar{\rho} + \bar{P}) - 3H\bar{P}_c -\left(\frac{\partial \bar{\rho}}{\partial \bar{n}}\right)_T(\bar{n}\bar{\Gamma}-3H\bar{n})\,.$$
The above equation can also be written as:
\begin{eqnarray}
  \dot{\bar{T}} = -\frac{1}{({\partial \bar{\rho}}/{\partial \bar{T}})_n}\left\{\left[(\bar{\rho} + \bar{P})-\bar{n}\left(\frac{\partial \bar{\rho}}{\partial \bar{n}}\right)_T \right]3H + 3H\bar{P}_c + \left(\frac{\partial \bar{\rho}}{\partial \bar{n}}\right)_T \bar{n}{\Gamma} \right\}\,.
\label{tdit}
\end{eqnarray}
To proceed further we have to spend some time on thermodynamic relations.
From Eq.~(\ref{1stl}) (for the background cosmological model) we can write
$$d\bar{s} = \frac{1}{\bar{n}\bar{T}}\left[d\bar{\rho} - \left(\frac{\bar{\rho} + \bar{P}}{\bar{n}}\right)d\bar{n}\right]\,,$$
which gives,
\begin{eqnarray}
\left(\frac{\partial \bar{s}}{\partial \bar{n}}\right)_T &=& \frac{1}{\bar{n}\bar{T}}\left[\left(\frac{\partial \bar{\rho}}{\partial \bar{n}}\right)_T - \left(\frac{\bar{\rho} + \bar{P}}{\bar{n}}\right)\right]\,,
\label{dsdnt}\\
\left(\frac{\partial \bar{s}}{\partial \bar{T}}\right)_n &=& \frac{1}{\bar{n}\bar{T}}  \left(\frac{\partial \bar{\rho}}{\partial \bar{T}}\right)_n\,.
\label{dsdtn}
\end{eqnarray}
For partial derivatives we know
$$\frac{\partial}{ \partial \bar{T}}\left\{\left(\frac{\partial \bar{s}}{\partial \bar{n}}\right)_T\right\}_n
=\frac{\partial}{ \partial \bar{n}}\left\{\left(\frac{\partial \bar{s}}{\partial \bar{T}}\right)_n\right\}_T\,,$$
which gives
$$\frac{\partial}{ \partial \bar{T}}\left\{\frac{1}{\bar{n}\bar{T}}\left[\left(\frac{\partial \bar{\rho}}{\partial \bar{n}}\right)_T - \left(\frac{\bar{\rho} + \bar{P}}{\bar{n}}\right)\right]\right\}_n
=\frac{\partial}{\partial \bar{n}}\left\{\frac{1}{\bar{n}\bar{T}}  \left(\frac{\partial \bar{\rho}}{\partial \bar{T}}\right)_n \right\}_T\,.$$
Calculating the derivatives in both sides and cancelling appropriate terms the above equation yields \cite{Weinberg:1971mx}
\begin{eqnarray}
\bar{T}\left(\frac{\partial \bar{P}}{\partial \bar{T}}\right)_n=(\bar{\rho} + \bar{P}) -\bar{n}\left(\frac{\partial \bar{\rho}}{\partial \bar{n}}\right)_T\,.
\label{therm1}
\end{eqnarray}
Using this relation in Eq.~(\ref{tdit}) we get
$$\dot{\bar{T}} = -\frac{1}{({\partial \bar{\rho}}/{\partial \bar{T}})_n}\left\{3H\bar{T}\left(\frac{\partial \bar{P}}{\partial \bar{T}}\right)_n + 3H\bar{P}_c + \left(\frac{\partial \bar{\rho}}{\partial \bar{n}}\right)_T
\bar{n}{\Gamma} \right\}\,,$$
which can also be written as
\begin{eqnarray}
\frac{\dot{\bar{T}}}{\bar{T}}=-3H \left(\frac{\partial \bar{P}}{\partial \bar{\rho}}\right)_n
-\frac{3H\bar{P}_c + ({\partial \bar{\rho}}/{\partial \bar{n}})_T \,\,\bar{n}{\Gamma}}{\bar{T}({\partial \bar{\rho}}/{\partial \bar{T}})_n}\,,
\label{tl1}
\end{eqnarray}
where we have used $$\frac{({\partial \bar{P}}/{\partial \bar{T}})_n}{({\partial \bar{\rho}}/{\partial \bar{T}})_n}=\left(\frac{\partial \bar{P}}{\partial \bar{\rho}}\right)_n\,.$$
If specific entropy is conserved then the expression of $\bar{P}_c$ is given as in Eq.~(\ref{crepi}) and we know the expression of $\bar{n}({\partial \bar{\rho}}/{\partial \bar{n}})_T$ from Eq.~(\ref{therm1}). Using these expressions we can write
$$3H\bar{P}_c + \bar{n}\left(\frac{\partial \bar{\rho}}{\partial \bar{n}}\right)_T {\Gamma} =
-{\Gamma}(\bar{\rho} + \bar{P})
+\left[(\bar{\rho} + \bar{P}) -\bar{T}\left(\frac{\partial \bar{P}}{\partial \bar{T}}\right)_n\right]\Gamma
=-\bar{T}\,\,{\Gamma}\left(\frac{\partial \bar{P}}{\partial \bar{T}}\right)_n\,.$$
Using this result in Eq.~(\ref{tl1}) for cosmological evolution conserving specific entropy we get
$$\frac{\dot{\bar{T}}}{\bar{T}}=-3H \left(\frac{\partial \bar{P}}{\partial \bar{\rho}}
\right)_n + {\Gamma} \left(\frac{\partial \bar{P}}{\partial \bar{\rho}}\right)_n\,,$$
and using the expression of $\dot{n}/n$ the above equation becomes
\begin{eqnarray}
\frac{\dot{\bar{T}}}{\bar{T}}=\left(\frac{\partial \bar{P}}{\partial \bar{\rho}}\right)_n \frac{\dot{\bar{n}}}{\bar{n}}\,. 
\label{tl2}
\end{eqnarray}
\vskip 1cm
%%%%%%%%%%%%%%%%%%%%%%%%%%%%%%%%%%%%%%%%%%%%%%%%
\begin{center}
{\bf Data availability statement}
\end{center}
Data sharing not applicable to this article as no datasets were generated or analysed during the current study.
%%%%%%%%%%%%%%%%%%%%%%%%%%%%%%%%%%%%%%%%%%%%%%%%%%%%%%%%%%%%%%%%%%%%%%%%%%%

%%%%%%%%%%%%%%%%%%%%%%%%%

\begin{thebibliography}{99}

%\cite{Berera:1995ie}
\bibitem{Berera:1995ie}
A.~Berera,
%``Warm inflation,''
Phys. Rev. Lett. \textbf{75}, 3218-3221 (1995)
doi:10.1103/PhysRevLett.75.3218
[arXiv:astro-ph/9509049 [astro-ph]].

%\cite{Hall:2003zp}
\bibitem{Hall:2003zp}
L.~M.~Hall, I.~G.~Moss and A.~Berera,
%``Scalar perturbation spectra from warm inflation,''
Phys. Rev. D \textbf{69}, 083525 (2004)
doi:10.1103/PhysRevD.69.083525
[arXiv:astro-ph/0305015 [astro-ph]].

%\cite{Matsuda:2009eq}
\bibitem{Matsuda:2009eq}
T.~Matsuda,
%``Evolution of the curvature perturbations during warm inflation,''
JCAP \textbf{06}, 002 (2009)
doi:10.1088/1475-7516/2009/06/002
[arXiv:0905.0308 [astro-ph.CO]].
%18 citations counted in INSPIRE as of 29 Jun 2020

%\cite{Visinelli:2014qla}
\bibitem{Visinelli:2014qla}
L.~Visinelli,
%``Cosmological perturbations for an inflaton field coupled to radiation,''
JCAP \textbf{01}, 005 (2015)
doi:10.1088/1475-7516/2015/01/005
[arXiv:1410.1187 [astro-ph.CO]].
%21 citations counted in INSPIRE as of 29 Jun 2020

%\cite{Parker:1969au}
\bibitem{Parker:1969au}
L.~Parker,
%``Quantized fields and particle creation in expanding universes. 1.,''
Phys. Rev. \textbf{183}, 1057-1068 (1969)
doi:10.1103/PhysRev.183.1057

\cite{Ford:1986sy}
\bibitem{Ford:1986sy}
L.~H.~Ford,
%``Gravitational Particle Creation and Inflation,''
Phys. Rev. D \textbf{35}, 2955 (1987)
doi:10.1103/PhysRevD.35.2955
%395 citations counted in INSPIRE as of 07 Dec 2020

%\cite{Lima:2012cm}
\bibitem{Lima:2012cm} 
  J.~A.~S.~Lima, S.~Basilakos and F.~E.~M.~Costa,
  %``New Cosmic Accelerating Scenario without Dark Energy,''
  Phys.\ Rev.\ D {\bf 86}, 103534 (2012)
  doi:10.1103/PhysRevD.86.103534
  [arXiv:1205.0868 [astro-ph.CO]].
  %%CITATION = doi:10.1103/PhysRevD.86.103534;%%
  %57 citations counted in INSPIRE as of 17 Nov 2019


%\cite{Gunzig:1997tk}
\bibitem{Gunzig:1997tk}
E.~Gunzig, R.~Maartens and A.~V.~Nesteruk,
%``Inflationary cosmology and thermodynamics,''
Class. Quant. Grav. \textbf{15}, 923-932 (1998)
doi:10.1088/0264-9381/15/4/014
[arXiv:astro-ph/9703137 [astro-ph]].
%70 citations counted in INSPIRE as of 29 Jun 2020


%\cite{Abramo:1996ip}
\bibitem{Abramo:1996ip}
L.~Abramo and J.~Lima,
%``Inflationary models driven by adiabatic matter creation,''
Class. Quant. Grav. \textbf{13}, 2953-2964 (1996)
doi:10.1088/0264-9381/13/11/011
[arXiv:gr-qc/9606064 [gr-qc]].
%67 citations counted in INSPIRE as of 29 Jun 2020

%\cite{Prigogine:1989zz}
\bibitem{Prigogine:1989zz}
I.~Prigogine, J.~Geheniau, E.~Gunzig and P.~Nardone,
%``Thermodynamics and cosmology,''
Gen. Rel. Grav. \textbf{21}, 767-776 (1989)
doi:10.1007/BF00758981
%215 citations counted in INSPIRE as of 29 Jun 2020

%\cite{Alcaniz:1999hu}
\bibitem{Alcaniz:1999hu}
J.~S.~Alcaniz and J.~A.~S.~Lima,
%``Closed and open FRW cosmologies with matter creation: Kinematic tests,''
Astron. Astrophys. \textbf{349}, 729-734 (1999)
[arXiv:astro-ph/9906410 [astro-ph]].
%38 citations counted in INSPIRE as of 30 Jun 2020

%\cite{Wichoski:1997hx}
\bibitem{Wichoski:1997hx}
U.~F.~Wichoski and J.~A.~S.~Lima,
%``Big bang cosmology with photon creation,''
Phys. Lett. A \textbf{262}, 103-109 (1999)
doi:10.1016/S0375-9601(99)00681-7
[arXiv:astro-ph/9708215 [astro-ph]].
%6 citations counted in INSPIRE as of 30 Jun 2020

%\cite{Pan:2018ibu}
\bibitem{Pan:2018ibu}
S.~Pan, J.~D.~Barrow and A.~Paliathanasis,
%``Two-fluid solutions of particle-creation cosmologies,''
Eur. Phys. J. C \textbf{79}, no.2, 115 (2019)
doi:10.1140/epjc/s10052-019-6627-5
[arXiv:1812.05493 [gr-qc]].
%7 citations counted in INSPIRE as of 30 Jun 2020

%\cite{Paliathanasis:2016dhu}
\bibitem{Paliathanasis:2016dhu}
A.~Paliathanasis, J.~D.~Barrow and S.~Pan,
%``Cosmological solutions with gravitational particle production and nonzero curvature,''
Phys. Rev. D \textbf{95}, no.10, 103516 (2017)
doi:10.1103/PhysRevD.95.103516
[arXiv:1610.02893 [gr-qc]].
%18 citations counted in INSPIRE as of 30 Jun 2020

%\cite{Lima:2014hda}
\bibitem{Lima:2014hda}
J.~A.~S.~Lima and I.~Baranov,
%``Gravitationally Induced Particle Production: Thermodynamics and Kinetic Theory,''
Phys. Rev. D \textbf{90}, no.4, 043515 (2014)
doi:10.1103/PhysRevD.90.043515
[arXiv:1411.6589 [gr-qc]].
%40 citations counted in INSPIRE as of 30 Jun 2020

%\cite{Baranov:2019qbc}
\bibitem{Baranov:2019qbc}
I.~P.~R.~Baranov, J.~F.~Jesus and J.~A.~S.~Lima,
%``Testing creation cold dark matter cosmology with theradiation temperature-redshift relation,''
Gen. Rel. Grav. \textbf{51}, no.2, 33 (2019)
doi:10.1007/s10714-019-2516-3
[arXiv:1605.04857 [astro-ph.CO]].
%6 citations counted in INSPIRE as of 30 Jun 2020

%\cite{Lima:2015xpa}
\bibitem{Lima:2015xpa}
J.~A.~S.~Lima, R.~C.~Santos and J.~V.~Cunha,
%``Is $\Lambda$CDM an effective CCDM cosmology?,''
JCAP \textbf{03}, 027 (2016)
doi:10.1088/1475-7516/2016/03/027
[arXiv:1508.07263 [gr-qc]].
%14 citations counted in INSPIRE as of 30 Jun 2020

%\cite{Steigman:2008bc}
\bibitem{Steigman:2008bc}
G.~Steigman, R.~C.~Santos and J.~A.~S.~Lima,
%``An Accelerating Cosmology Without Dark Energy,''
JCAP \textbf{06}, 033 (2009)
doi:10.1088/1475-7516/2009/06/033
[arXiv:0812.3912 [astro-ph]].
%57 citations counted in INSPIRE as of 30 Jun 2020

%\cite{Guth:1980zm}
\bibitem{Guth:1980zm}
A.~H.~Guth,
%``The Inflationary Universe: A Possible Solution to the Horizon and Flatness Problems,''
Adv. Ser. Astrophys. Cosmol. \textbf{3}, 139-148 (1987)
doi:10.1103/PhysRevD.23.347
%7894 citations counted in INSPIRE as of 30 Jun 2020

%\cite{Kolb:1993rv}
\bibitem{Kolb:1993rv}
E.~W.~Kolb,
%``The Inflationary decade,''
Phys. Rept. \textbf{227}, 5-12 (1993)
doi:10.1016/0370-1573(93)90052-F
%4 citations counted in INSPIRE as of 30 Jun 2020

%\cite{Linde:1983gd}
\bibitem{Linde:1983gd}
A.~D.~Linde,
%``Chaotic Inflation,''
Phys. Lett. B \textbf{129}, 177-181 (1983)
doi:10.1016/0370-2693(83)90837-7
%2918 citations counted in INSPIRE as of 30 Jun 2020

%\cite{Zimdahl:1993yu}
\bibitem{Zimdahl:1993yu}
W.~Zimdahl, D.~Pavon and D.~Jou,
%``Cosmological perturbations in a universe with particle production,''
Class. Quant. Grav. \textbf{10}, 1775-1789 (1993)
doi:10.1088/0264-9381/10/9/019
%8 citations counted in INSPIRE as of 07 Dec 2020

%\cite{Calvao:1991wg}
\bibitem{Calvao:1991wg} 
M.~O.~Calvao, J.~A.~S.~Lima and I.~Waga,
%``On the thermodynamics of matter creation in cosmology,''
Phys.\ Lett.\ A {\bf 162}, 223 (1992).
doi:10.1016/0375-9601(92)90437-Q
%%CITATION = doi:10.1016/0375-9601(92)90437-Q;%%
%181 citations counted in INSPIRE as of 17 Nov 2019

%\cite{Pan:2016jli}
\bibitem{Pan:2016jli} 
  S.~Pan, J.~de Haro, A.~Paliathanasis and R.~J.~Slagter,
  %``Evolution and Dynamics of a Matter creation model,''
  Mon.\ Not.\ Roy.\ Astron.\ Soc.\  {\bf 460}, no. 2, 1445 (2016)
  doi:10.1093/mnras/stw1034
  [arXiv:1601.03955 [gr-qc]].
  %%CITATION = doi:10.1093/mnras/stw1034;%%
  %23 citations counted in INSPIRE as of 17 Nov 2019

%\cite{Lima:1995xz}
\bibitem{Lima:1995xz} 
  J.~A.~S.~Lima, A.~S.~M.~Germano and L.~R.~W.~Abramo,
  %``FRW type cosmologies with adiabatic matter creation,''
  Phys.\ Rev.\ D {\bf 53}, 4287 (1996)
  doi:10.1103/PhysRevD.53.4287
  [gr-qc/9511006].
  %%CITATION = doi:10.1103/PhysRevD.53.4287;%%
  %134 citations counted in INSPIRE as of 17 Nov 2019
  
%\cite{Weinberg:1971mx}
\bibitem{Weinberg:1971mx} 
  S.~Weinberg,
  %``Entropy generation and the survival of protogalaxies in an expanding universe,''
  Astrophys.\ J.\  {\bf 168}, 175 (1971).
  doi:10.1086/151073
  %%CITATION = doi:10.1086/151073;%%
  %439 citations counted in INSPIRE as of 18 Nov 2019

%\cite{Ford:1986sy}
%\bibitem{Ford:1986sy}
%L.~H.~Ford,
%``Gravitational Particle Creation and Inflation,''
%Phys. Rev. D \textbf{35}, 2955 (1987)
%doi:10.1103/PhysRevD.35.2955
%395 citations counted in INSPIRE as of 07 Dec 2020

%\cite{Chakraborty:2014fia}
\bibitem{Chakraborty:2014fia}
S.~Chakraborty, S.~Pan and S.~Saha,
%``A third alternative to explain recent observations: Future deceleration,''
Phys. Lett. B \textbf{738}, 424-427 (2014)
doi:10.1016/j.physletb.2014.10.009
[arXiv:1411.0941 [gr-qc]].
%39 citations counted in INSPIRE as of 07 Dec 2020
  
%\cite{Mukhanov:1990me}
\bibitem{Mukhanov:1990me} 
  V.~F.~Mukhanov, H.~A.~Feldman and R.~H.~Brandenberger,
  %``Theory of cosmological perturbations. Part 1. Classical perturbations. Part 2. Quantum theory of perturbations. Part 3. Extensions,''
  Phys.\ Rept.\  {\bf 215}, 203 (1992).
  doi:10.1016/0370-1573(92)90044-Z
  %%CITATION = doi:10.1016/0370-1573(92)90044-Z;%%
  %2859 citations counted in INSPIRE as of 25 Jan 2020

%\cite{Domenech:2021ztg}
\bibitem{Domenech:2021ztg}
G.~Dom\`enech,
%``Scalar Induced Gravitational Waves Review,''
Universe \textbf{7}, no.11, 398 (2021)
doi:10.3390/universe7110398
[arXiv:2109.01398 [gr-qc]].
%27 citations counted in INSPIRE as of 23 Jan 2022

%\cite{Carroll:2003st}
\bibitem{Carroll:2003st}
S.~M.~Carroll, M.~Hoffman and M.~Trodden,
%``Can the dark energy equation-of-state parameter $w$ be less than $-1$?,''
Phys. Rev. D \textbf{68}, 023509 (2003)
doi:10.1103/PhysRevD.68.023509
[arXiv:astro-ph/0301273 [astro-ph]].
%1080 citations counted in INSPIRE as of 25 Jan 2022
%%%%%%%%%%%%%%%%%%%%%%%%%%%%%%%%%%%%%%%%%%%%%%%%
\end{thebibliography}
\end{document}